\documentclass[12pt]{iopart}

\usepackage{epsfig}
\usepackage{graphicx,color}

\usepackage{iopams}

\usepackage{graphicx,color}
\usepackage{epsfig}

\usepackage{dcolumn}

\usepackage{bm}
\usepackage{amssymb}
\usepackage{array}

    \newcommand{\ket}[1]{\ensuremath{|\,{#1}\,\rangle}}

    \newcommand{\itg}[1]{\ensuremath{\int\!\!d{#1}\!\!}}
    \newcommand{\itgf}[1]{\ensuremath{\int\!\!d{#1}\,}}

    \newcommand{\sinc}{\ensuremath{\mbox{\hspace{1.3pt}sinc}\,}}

\begin{document}

\title{Controlling the transverse correlation in QPM parametric down-conversion}

\author{O. Cosme,$^{1}$ A. Delgado,$^{2}$ G. Lima,$^{2,*}$ C. H. Monken,$^{1}$ and S. P\'{a}dua$^{1}$}

\address{
$^1$Departamento de F\'{\i}sica,
Universidade Federal de Minas Gerais, Caixa Postal 702, Belo~Horizonte, MG 30123-970, Brazil. \\
$^2$Center for Quantum Optics and Quantum Information,
Departamento de F\'{\i}sica, Universidad de Concepci\'on,  Casilla 160-C, Concepci\'on, Chile. \\
$^*$Corresponding author: glima@udec.cl}

\begin{abstract}
In this work we study the transverse spatial correlation of the pair of photons generated via the process of spontaneous parametric frequency down-conversion, in periodically poled non-linear crystals illuminated by a pulsed laser beam. It is well known that the two-photon state generated in quasi-phase matching (QPM) configurations depends explicitly on the characteristics of the pump beam, on the crystal modulated non-linearity, and on the detection geometry. This has allowed the development of several techniques for controlling the biphoton spectral and spatial properties. Here we discuss another technique for implementing the spatial entanglement modification in QPM gratings. We show, theoretically and experimentally, that in nearly collinear geometries, the spatial shape modulation of the pump beam allows for the control of the biphoton transverse spatial correlation.
\end{abstract}

\pacs{03.65.-w, 03.67.Mn, 42.65.Lm}

\submitto{\jpb}
\maketitle

\section{Introduction}

In the process of spontaneous parametric down-conversion (SPDC), one photon from a pump beam incident onto a nonlinear birefringent crystal splits, with small probability, into two lower-frequency photons usually called signal and idler or also biphoton. Due to its flexibility, such as the efficient three-wave mixing at user-selectable wavelengths, and the controllable degree of entanglement of the down-converted photons, the SPDC has been used as the main light source for quantum information and many others quantum optics applications \cite{Bouwmeester}.

The state of the biphoton generated in this process, can be obtained from the Hamiltonian describing the coupling of the incident pump field to the down-converted fields over the region of the non-linear medium \cite{Mandel,Klyshko}. For a thin non-linear rectangular crystal, pumped by a continuous (CW) laser beam operating with a frequency, $\omega_0$, and whose cross section is entirely contained in the crystal, the biphoton spatial and spectral properties may be written, considering a first-order perturbation description, as \cite{Rubin,Kulik,Monken,Fedorov,Monken2}

\begin{eqnarray} \label{Rubin}
\ket{\Psi} &\propto &\itg{\nu} \itg{\mathbf{q}_s} \itgf{\mathbf{q}_i} \tilde{E}(\mathbf{q}_s+ \mathbf{q}_i) \sinc{\left(\frac{\Delta \kappa_z L_z}{2}\right)} \nonumber \\
& &\times G_w(\nu) \exp{\left(-i \frac{\Delta \kappa_z L_z}{2}\right)} \ket{\mathbf{q}_s,\nu} \ket{\mathbf{q}_i,-\nu},
\end{eqnarray} where we have assumed that the crystal is centered at the origin and the pump beam propagating through the longitudinal $z$-direction. The state $\ket{\mathbf{q}_j,-\nu}$ represents one photon in mode $j$ ($j=i,s$) with transverse wave vector $\mathbf{q}_j$ and frequency $\frac{(1-\nu)\omega_0}{2}$. $L_z$ is the length of the crystal and $\Delta \kappa_z$ the phase mismatch along the longitudinal direction. $\tilde{E}$ is the angular spectrum  of the incident pump filed, and $G_w$ is the the frequency distribution of the down-converted fields, which is dependent on the interference filters placed in the propagation path of the down-converted beams. The phase matching condition, dictated by the sinc function, can be satisfied in non-linear crystals just for certain polarization modes \cite{Burnham}.

From Eq.~(\ref{Rubin}), one can see that the two-photon state carries information about the initial pump beam. This transfer of information allows, therefore, the control of the joint properties of the down-converted fields. This has motivated the development of several techniques for preparing specific two-photon states. Concerning their spectral properties, for example, it has been shown
that in non-collinear geometries, the pump beam modulation can be used to shape adequately the biphoton joint spectrum \cite{Carrasco}, and also that the application of angular dispersion techniques, to the incident pump beam, allows one to generate frequency-correlated and frequency-uncorrelated pairs of photons in the SPDC \cite{Torres}.

The dependence of the signal and idler transverse spatial correlation, on the pump beam properties, was studied in \cite{Monken}. The authors showed that for thin non-linear crystals, the transverse probability of coincidence detection, in nearly collinear geometries, can be controlled exclusively through the manipulation of the spatial profile of the incident pump beam. This type of control found important applications and has been used for performing quantum imaging \cite{Ivan}, quantum interference \cite{Shih} and state engineering {\cite{Us,Steve}.

In this work we show that this technique of quantum control can also be employed for the down-converted fields generated in periodically poled crystals. The control of the biphoton states generated through down-conversion, in quasi-phase matching (QPM) gratings, has already been the subject of some experimental investigations \cite{Saleh,Uren,Shapiro,Ming,Exter1,Exter2}. In \cite{Saleh}, ultra-broadband biphotons were generated while considering non-linear gratings with linearly chirped poling periods, and in \cite{Shapiro}, the authors implemented a scheme for the generation of polarization maximally entangled states with periodically poled crystals. Another interesting experimental result was the demonstration, that one can use the transverse modulation of the non-linearity of these crystals, to control the transverse correlations of the down-converted fields \cite{Ming}. The manipulation of the spatial correlation of the down-converted photons, can also be done with optical systems mounted behind the non-linear crystal, as it was demonstrated in \cite{Exter2}. Our technique makes use of pump manipulation to control the biphoton spatial entanglement, and it can be used together with the transverse modulation of the crystal non-linearity, or the post-manipulation technique, for the generation of a broader class of spatially entangled two-photon states in the QPM down-conversion.

Because of the higher non-linear effective electric susceptibility of the periodically poled crystals, our results may have some relevant consequences for some of the aforementioned experiments, which explored the transverse correlation of the down-converted fields. In the field of quantum information, for example, the possibility of generating at higher rates, the entangled high-dimensional quantum systems reported in \cite{Us}, may allow their use for doing quantum communication \cite{Gisin}, or to experimentally test some Bell inequalities without the locality-loophole \cite{Bell}.

\section{The two-photon state generated in periodically poled crystals}

The two-photon state generated by SPDC in periodically poled crystals illuminated by CW laser beams can also be cast in a form similar to Eq.~(\ref{Rubin}), as it was showed in \cite{Carrasco}, while studying the dependence of the biphoton spectrum on the detection geometry and pump modulation. Here we extend the calculation of this two-photon state by considering a more general situation, where the incident laser is a pulsed beam. After this calculation we show, theoretically and experimentally, that the control via pump manipulation, of the transverse spatial correlations of the down-converted fields generated in non-linear periodic poled crystals, can be realized for nearly collinear configurations.

\subsection{Interaction Hamiltonian}

The interaction Hamiltonian can be written as \cite{Mandel,Klyshko}

\begin{equation} \label{Ham}
H_1=\frac{1}{2} \int_{V}d\mathbf{r} \, \chi(\mathbf{r})^{(2)} E_p(\mathbf{r},t)E_s(\mathbf{r},t)E_i(\mathbf{r},t),
\end{equation} where the integration is taken over the volume $V$ of the non-linear medium. $\chi^{(2)}$ is the non-linear electric susceptibility. In our calculations we consider that $\chi^{(2)}$ is wavelength-independent as is commonly assumed \cite{Saleh,Ming} \footnote{ This is equivalent to say that we are restricting our calculations to the state of the down-converted photons whose frequency-spread is small compared to their central frequency $\frac{w_{0}}{2}$. This is exactly the case of a typical experiment where narrow interference filters with a FWHM of $5$nm are used in front of the detectors. It is worth to mention that under such simplification, the functions $l_i[\omega_i(\mathbf{k_i})]$ (and $l_s[\omega_s(\mathbf{k_s})]$) of Eq.~(\ref{Equant}), can be regarded as constants.}. Due to its intensity, the pump electric field, $E_p$, is taken to be classical, while signal and idler fields ($E_s$ and $E_i$) are quantized. We also consider that they have well defined polarizations \footnote{We will consider that a type-II crystal is illuminated by a horizontally polarized pump beam and for simplifying the analysis, we also consider that the two-photon state being calculated here is composed of the idler photon with vertical polarization and the signal photon with the horizontal polarization. We stress that in this work we are not interested in the polarization entanglement of the down-converted fields, which can be generated in this type of process when one properly considers the longitudinal walk-off effects of the type-II down-conversion \cite{Shapiro}.}. First we assume that the pump beam is propagating through the longitudinal $z$-direction, with a well defined central frequency, $\omega_0$, and that the pulse shape modulation of the beam can be approximated by a temporal Gaussian function, so that

\begin{eqnarray} \label{Pump}
E_p(\mathbf{r},t) &= &e^{-\Gamma \left(t-
\left(\frac{n_{g}}{c}z\right)\right)^2}e^{-i\omega_{0}
t} \times \itg{k_{0x}} \itgf{k_{0y}}\tilde{E}(\mathbf{q_{0}})e^{i \mathbf{k_0}.\mathbf{r}},
\end{eqnarray} where $\mathbf{k_0}$ represents the wave number of the pump beam and $\mathbf{q_0}$ its transverse component (so that $\mathbf{k_0}=\mathbf{k_{0z}}+\mathbf{q_0}$ and $\mathbf{q_0}=k_{0x}\mathbf{i}+k_{0y}\mathbf{j}$). The parameter $\Gamma$ is equal to the inverse of the square of the temporal pulse width and $n_g$ is the pulse group velocity refractive index. $\tilde{E}$ is the angular spectrum of the pump beam.

The positive-frequency part of the quantized signal field operator is written as \cite{Mandel}

\begin{equation} \label{Equant}
E_s^{(+)}(\mathbf{r},t)=\frac{1}{L^{\frac{3}{2}}}\sum_{\mathbf{k_s}} l_s[\omega_s(\mathbf{k_s})] a_{\mathbf{k_s}}(t) e^{i[\mathbf{k_s}.\mathbf{r}-\omega_s(\mathbf{k_s})t]},
\end{equation} where $l_s[\omega_s(\mathbf{k_s})]=\sqrt{\frac{\hbar \omega_s(\mathbf{k_s})}{2 \epsilon_0}}$, with $L^3$ being the quantization volume. $a_{\mathbf{k_s}}(t)$ represents the photon annihilation operator for the mode $\mathbf{k_s}$ and frequency $\omega_s(\mathbf{k_s})$. The idler field operator positive-frequency part, $E_i^{(+)}(\mathbf{r},t)$, is defined in an analogous form.

The longitudinal modulation of the non-linear electric susceptibility of a QPM grating with a periodic poling period, can be written in terms of a Fourier series as

\begin{equation} \label{chi}
\chi^{(2)}(z) \propto d_{eff} \sum_{m=-\infty}^{\infty}G_{m}e^{-i K_{m}z},
\end{equation} where $K_m$ is the spatial varying frequency, which is related to the modulation period, $\Lambda$, by $K_m=\frac{2 m \pi}{\Lambda}$. The coefficient $G_m$ is given in terms of the duty cycle of the crystal, $D$, by $G_m=\sinc(m \pi D)$. $d_{eff}$ is the effective non-linear coefficient of the crystal, which is a fixed material property \cite{Exter1}.

\subsection{The two-photon state}

Within a first order perturbation approach in the interaction picture, the two-photon state vector at the output of the non-linear crystal can be written as \cite{Rubin}

\begin{equation} \label{St1}
\ket{\Psi}=\ket{vac}-\frac{i}{\hbar}\int_0^{\tau} H_1(t) dt \ket{vac},
\end{equation} where $\ket{vac}$ is the vacuum state and $\tau$ is the interaction time.

If we assume that the cross section of the incident beam is smaller than the transversal dimensions of the non-linear crystal, we can calculate the integrals over the transverse coordinates ($x$,$y$) which appear in Eq.~(\ref{Ham}), with the integration limits extended to infinity. Therefore, the integrations which come from Eq.~(\ref{Ham}) and Eq.~(\ref{Pump}) may be simplified to

\begin{eqnarray} \label{Int2}
\itg{k_{0x}}\itg{k_{0y}} \int_{V}d\mathbf{r} \tilde{E}(\mathbf{q_{0}}) e^{i \mbox{\boldmath{$\Delta$}} \mbox{\boldmath{$\kappa$}} \cdot \mathbf{r}} e^{-\Gamma \left(t- \left(\frac{n_{g}}{c}z\right)\right)^2}  &  & \nonumber \\= \itg{k_{0x}}\itgf{k_{0y}}\delta(\Delta \kappa_x)\delta(\Delta \kappa_y)\tilde{E}(\mathbf{q_{0}}) \int_0^{L_z}dz e^{i \Delta \kappa_z z} e^{-\Gamma \left(t- \left(\frac{n_{g}}{c}z\right)\right)^2} & &\nonumber \\ = \tilde{E}(\mathbf{{q}_s} + \mathbf{{q}_i}) \int_0^{L_z}dz e^{i \Delta \kappa_z z} e^{-\Gamma \left(t- \left(\frac{n_{g}}{c}z\right)\right)^2}, & &
\end{eqnarray} where again, $\mathbf{q}_j$ denotes the transverse wave vector of the photon in mode $j$ ($j=i,s$) and $\mathbf{q_0}$ the  transverse component of the pump beam wave number, $\mathbf{k_0}$. One can clearly see the transverse momentum conservation present in the second line of Eq.~(\ref{Int2}). $L_z$ is the longitudinal size of the crystal and $\mbox{\boldmath{$\Delta$}}\mbox{\boldmath{$\kappa$}}$ is the vector associated with the different phase-matching conditions. For a longitudinal periodically poled non-linear crystal designed for m-order quasi-phase matching, the longitudinal phase mismatch, $\Delta \kappa_z$, is given by $\Delta \kappa_z = k_{0z} - k_{iz} - k_{sz} - \frac{2 \pi m}{\Lambda}$.

With the result of Eq.~(\ref{Int2}), we can now calculate the temporal integration of Eq.~(\ref{St1}). Since the crystal is fixed in a certain longitudinal position and the pump pulses are temporally limited, the integration limits can be extended to infinity and we get that

\begin{eqnarray} \int_{-\infty}^{+\infty} e^{-\Gamma \left(t- \left(\frac{n_{g}}{c}z\right)\right)^2} e^{-i(\omega_0-\omega_s-\omega_i)t}d t &= &e^{-i( n_{g}\Delta\omega/c)z}e^{-\frac{(\Delta\omega)^2}{2\Gamma}},
\end{eqnarray} where $\Delta\omega=(\omega_0-\omega_s-\omega_i)$.

Therefore, the integration over the longitudinal length of the crystal which enters in Eq.~(\ref{Int2}), will be given by

\begin{eqnarray}
\int_{0}^{L_{z}}dz
 e^{i(\Delta \kappa_z)z} e^{-i(n_{g} \Delta\omega/c)z}  &= &L_{z} e^{i \frac{L_{z}}{2}(\Delta \kappa_z - n_{g} \Delta\omega /c)} \sinc{\left[\frac{L_{z}}{2}(\Delta \kappa_z - n_{g} \Delta\omega /c)\right]}. \nonumber \\
\end{eqnarray}

Next we assume that the quantization volume of the down-converted fields is infinite \cite{Mandel}, and then we can write the two-photon state generated by SPDC, in a quasi-phase matching non-linear periodically poled crystal, illuminated by a pulsed beam, as

\begin{eqnarray} \label{Stfinal}
\ket{\Psi} &\propto &\itg{\omega_s} \itg{\omega_i} \itg{\mathbf{q}_s} \itgf{\mathbf{q}_i} \tilde{E}(\mathbf{q}_s+ \mathbf{q}_i) \sinc{\left[\frac{L_{z}}{2}(\Delta \kappa_z - n_{g} \Delta\omega /c)\right]} \nonumber \\
& &\times e^{-\frac{(\Delta\omega)^2}{2\Gamma}} e^{i \frac{L_{z}}{2}(\Delta \kappa_z - n_{g} \Delta\omega /c)} \ket{\mathbf{q}_s,\omega_s} \ket{\mathbf{q}_i,\omega_i}.
\end{eqnarray}

\subsection{Controlling the transverse correlation}

The transverse spatial correlation of the down converted beams generated in periodically poled crystals is therefore dependent on the angular spectrum of the incident beam and also on the QPM function, whose spatial profile is proportional to $\sinc^2{\left[\frac{L_{z}}{2}(\Delta \kappa_z - n_{g} \Delta\omega /c)\right]}$.

We can use the simple form of the paraxial approximation (which is valid for nearly collinear geometries and for configurations with negligible Poynting vector transverse walk-off, as is the case of periodically poled crystals) to rewrite $\Delta \kappa_z$ as

\begin{eqnarray} \label{Deltak}
\Delta \kappa_z &= &-k_{iz} - k_{sz} - \frac{2 \pi m}{\Lambda} + \sqrt{k_0^2-k_{0x}^2-k_{0y}^2} \nonumber \\
& &\approx k_0-k_{iz}-k_{sz}-\frac{2 \pi m}{\Lambda}-\frac{q_0^2}{2k_{0}} \nonumber \\
& &\approx \frac{1}{c}(n_{0}\omega_{0}-n_{i}\omega_{i}-n_{s}\omega_{s}) -\frac{2 \pi m}{\Lambda}+c\frac{q_{i}^2}{2n_{i}\omega_{i}}+c\frac{q_{s}^2}{2n_{s}\omega_{s}}-c\frac{|\mathbf{q}_{i}+\mathbf{q}_{s}|^2}{2n_{0}\omega_{0}}, \nonumber \\
\end{eqnarray} where $n_i$ ($n_s$) and $\omega_i$ ($\omega_s$) are the crystal refractive index and the angular frequency of the idler (signal) field, respectively. So, the function $A\equiv\Delta \kappa_z - n_{g} \frac{\Delta\omega}{c}$, will be given by

\begin{eqnarray}
A &= &\frac{ n_0 \omega_0 - n_i \omega_i - n_s \omega_s}{c} - \frac{n_g}{c}(\omega_0 - \omega_i - \omega_s) \nonumber \\
& &+\frac{1}{2c}\left(\sin^2(\alpha_i)n_{i} \omega_i+ \sin^2(\alpha_s)n_{s} \omega_s - \frac{(\sin(\alpha_i) n_{i} \omega_i- \sin(\alpha_s)n_{s} \omega_s)^2}{n_0 \omega_0}\right) \nonumber \\
& &- \frac{2\pi m}{\Lambda}, \label{AA}
\end{eqnarray} where we have assumed, without loss of generalization, that the transverse components of the idler and signal wave vectors ($\mathbf{q}_s$ and $\mathbf{q}_i$) lie along the direction-$x$. We also used that $\mathbf{q}_j=\mathbf{k}_j \sin{(\alpha_j)}$. The relation between these vectors and the angles $\alpha_i$ and $\alpha_s$ is shown in the inset of Fig.~\ref{Fig:Setup}.

Let us consider the down-converted beams with $\omega_i+\omega_s=\omega_0$, so that we can neglect the effects of the second term of Eq.~(\ref{AA}), related with $n_g$. In the paraxial approximation, we may assume that $\alpha_i \approx \alpha_s=\alpha$, and under such considerations Eq.~(\ref{AA}) simplifies to

\begin{eqnarray}
A &\approx &\frac{ n_0 \omega_0 - n_i \omega_i - n_s \omega_s}{c} \nonumber \\
& &+\frac{\sin^2(\alpha)}{2c}\left(n_{i} \omega_i+n_{s} \omega_s - \frac{(n_{i} \omega_i-n_{s} \omega_s)^2}{n_0 \omega_0}\right) \nonumber \\
& &- \frac{2 \pi m}{\Lambda}. \label{AA2}
\end{eqnarray}

In general, the poling period of a QPM crystal is done for collinear generation, which means that the third line of Eq.~(\ref{AA2}) is used to cancel out its first line. In this case, the transverse spatial phase-matching profile, which is also known as Maker fringes \cite{Exter1,Maker}, is determined by the second line of Eq.~(\ref{AA2}). The Maker fringes will have a maximum at its central ring and it happens when $\alpha$ is null. The efficiency of the down-conversion process will decrease with the increasing of the emission angle of the down-converted beams.

\begin{figure}[tbh]
\vspace{2.5cm}
\begin{center}
\includegraphics[width=0.55\textwidth]{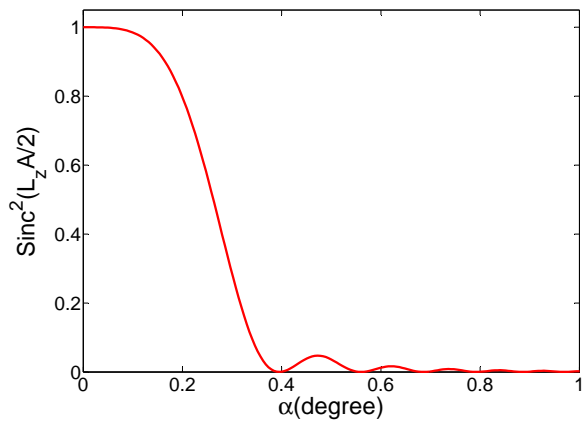}
\end{center}
\vspace{-2.5cm}
\caption{In this figure it is shown the QPM efficiency as a function of the emission angle $\alpha$. This curve corresponds to the spatial phase-matching profile which is also know as Maker fringes. See text for details.} \label{FigA}
\end{figure}

In Fig.~\ref{FigA}, we show the plot of the transverse spatial phase-matching profile. For obtaining this curve, we considered the refractive indexes corresponding to the degenerate fields created by type-II SPDC in periodically poled potassium titanyl phosphate (PPKTP) crystals. We considered a incident radiation of $413$~nm, polarized at the y-direction, and propagating through the crystallographic z axis. The down-converted fields have $x$- and $y$-polarizations. The crystal is periodically poled along its crystallographic $z$ axis. The associated first-order quasi-phase matching poling period is $\Lambda=11.4617$~$\mu$m. The crystal considered has the effective longitudinal length of $0.96$~cm. In this situation, one can see that for a detection solid angle of just $0.5^{\circ}$, the QPM down-conversion efficiency is already very low. Therefore, to be able to manipulate the spatial distribution of the down-converted beams via pump modulation, one must be able to detect them in configurations which are nearly collinear, where the $\sinc$ function will be close to $1$ and almost constant. This can be done by detecting the down-converted beams in the far field plane, and over small transverse distances. In this case, the emission angle of the down-converted beams are related with the detectors transverse positions by \cite{Monken2}

\begin{eqnarray}
\alpha_i \approx \frac{p_i}{z_D}, \\
\alpha_s \approx \frac{p_s }{z_D},
\end{eqnarray} where $p_j$ is the transverse position of the detector in mode $j$ with $j=i,s$, and $z_D$ is the longitudinal distance between the crystal and the detection plane. For example, for a distance $z_D$ of $1$~m, we have that over a transverse displacement of $3$~mm, there is a reduction in the QPM down-conversion efficiency that is smaller than $1\%$. So under this limit, the transverse spatial correlation of the down-converted beams will be dictated exclusively by the spatial shape modulation of the incident pump beam.

It is also worth to mention that the spatial profile of the pump beam, which will be used to manipulate the transverse correlation of the down-converted photons, must be of the order of these small transverse distances, that is, the spatial distribution of the pump beam, at the plane of detection, must be small when compared with the spatial phase-matching profile.

\section{Experimental setup and results}

The experimental apparatus represented in Fig.~\ref{Fig:Setup}, was carried out to demonstrate the feasibility of our technique for controlling the transverse spatial correlations of the down-converted beams generated in periodically poled crystals. A potassium titanyl phosphate crystal, periodically poled for collinear type-II parametric down-conversion luminescence, was pumped by a $100$~mW tunable frequency-doubled Ti:sapphire laser operating at $413$~nm, at the repetition rate of $76$~Mhz and with pulses of $200$~fs. The crystal was manufactured by Raicol Crystals Ltd. and it has an effective length of $9.6$~mm, and a cross area of $1$mm$\times$$1$mm. Its poling period is $\Lambda=11.4617$~$\mu$m. The crystal temperature was stabilized at $40^{\circ}$. Before reaching the crystal, the laser that is operating in a TEM$_{00}$ mode, passes though a telescope system which is used to reduce its radius at $e^{-2}$ of maximum irradiance, from $1$mm to $0.5$mm. After transmission by the telescope, the laser is collimated and its cross section is smaller than the cross section of the crystal. After being transmitted through the crystal, the pump beam is reflected out by a dichroic mirror (DM) and detected with a power meter. Down-converted photons with orthogonal polarizations and at the degenerated wavelength of $826$~nm, are selected by using interference filters with small bandwidths ($2$~nm FWHM). They are generated in nearly collinear spatial modes which are separated after their transmission through a polarizing beam splitter (PBS). They are then detected by avalanche photodiode detectors $D_i$ and $D_s$, connected to a circuit (C) used to record the singles and coincidences counts. In front of each detector and of the power meter, there is a single slit of width of $0.1$~mm which further restricts the spatial modes being detected. The detectors and the power meter were placed at a equal distance of $50$~cm from the non-linear crystal.

\begin{figure}[tbh]
\vspace{-2cm}
\begin{center}
\includegraphics[width=0.55\textwidth]{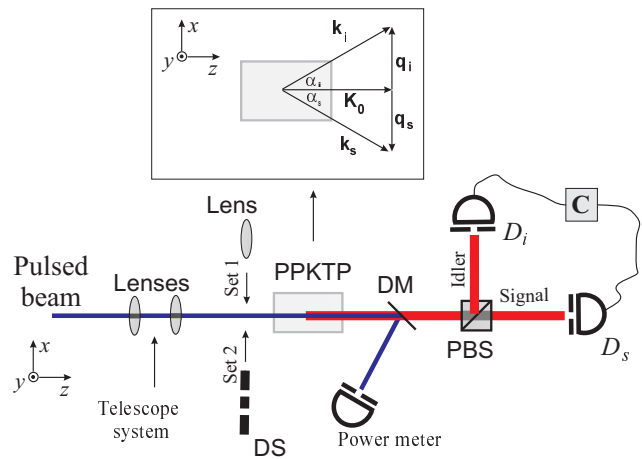}
\end{center}
\vspace{-3cm}
\caption{Experimental setup. In this figure, PPKTP represents the periodically poled crystal used in our experiment, DS the double slit, DM the dichroic mirror, and $D_i$, $D_s$ the avalanche photo detectors. PBS is a polarizing beam splitter and C is the circuit used to record the coincidences.} \label{Fig:Setup}
\end{figure}

As it was extensively discussed in \cite{Monken}, for experimental configurations where the transverse correlation of the down-converted fields are governed only by the pump beam angular spectrum, the coincidences counts recorded as a function of the the detectors transverse positions, $C(\mathbf{p}_s,\mathbf{p}_i)$, will be proportional to the intensity profile of the pump beam at the detection plane. That is, $C(\mathbf{p}_s,\mathbf{p}_i) \propto |W\mathbf{(R)}|^2$, where $W$ is the pump beam amplitude profile and $\mathbf{R}$, a weighted mean of $\mathbf{p_s}$ and $\mathbf{p_i}$. When both detectors are scanned in the same transverse direction (lets say the $x$-direction) and with the same step, the coincidences counts will have the same shape of the pump beam intensity profile.

To show this transfer of information for the states generated in our experiment, we performed two distinct sets of measurements. In the first one, a lens was placed in the propagation path of the pump beam (See Fig.~\ref{Fig:Setup}), so that its cross section was slightly enlarged at the detection plane. The lens was placed at a distance of approximately $3$~cm in front of the crystal, and it had a focal length of $50$~cm. The pump beam transverse intensity profile was measured with the power meter, and the coincidence counts were recorded as a function of the detectors $D_i$ and $D_s$ transverse displacement. They were moved in the same direction and with the same step. The experimental results obtained for the pump intensity profile and the coincidences recorded are shown in Fig.~\ref{Fig:Results}(a) and Fig.~\ref{Fig:Results}(b), respectively. One can clearly see the good agrement between the distribution of these curves. In this configuration we have that the spatial profile of the pump beam is much smaller than of the QPM spatial profile, and that the total transverse displacement of the detectors corresponds to a reduction in the QPM efficiency which is smaller than $1\%$.

In the second experimental configuration, the lens used in the first configuration was removed and one double slit, with its slits width of $100$~$\mu m$ and which were separated by $200$~$\mu m$, was introduced in the propagation path of the pump beam at a distance of $1$~cm in front of the crystal. This double slit produces Young interference fringes in the plane of the detectors. The interference pattern recorded with the power meter is shown in Fig.~\ref{Fig:Results}(c). The corresponding coincidences counts, recorded in the same way described above, are shown in Fig.~\ref{Fig:Results}(d). Again it is clear the good agrement between these curves. Therefore, these results demonstrate, for nearly collinear geometries, that the transverse correlation of the biphotons generated by quasi-phase-matched down-conversion can indeed be controlled through the pump spatial shape modulation.

\begin{figure}[tbh]
\vspace{3.5cm}
\begin{center}
\includegraphics[width=0.65\textwidth]{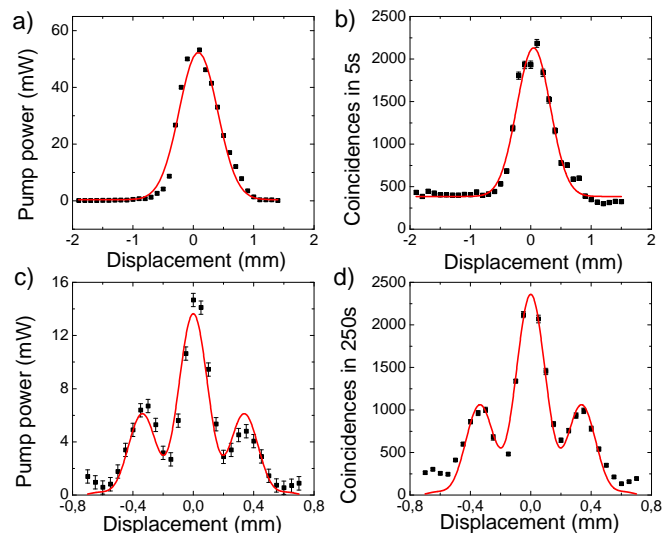}
\end{center}
\vspace{-2.5cm}
\caption{Experimental results. In (a) it is shown the pump intensity profile for the first experimental configuration. In (b), it is shown the coincidences recorded as function of the transverse position of the detectors $D_i$ and $D_s$, for the same configuration of (a). In (c) it is shown the pump intensity profile for the second experimental configuration and in (d), the corresponding coincidences.} \label{Fig:Results}
\end{figure}

\section{Conclusion}

In this work we have studied the control of the transverse spatial correlation of the biphotons generated through spontaneous frequency parametric down-conversion in periodically poled non-linear crystals. We have demonstrated, theoretically and experimentally, that for nearly collinear experimental configurations, this control can be done via the modulation of the spatial shape of the incident pump beam. Because of the higher non-linear effective electric susceptibility of the periodically poled crystals, this result which demonstrates the transfer of the angular spectrum shape of the incident pump beam to the biphoton wave function at the QPM down-conversion, may have some relevant consequences for the experiments which explored the transverse correlation of the down-converted fields. In the field of quantum information, for example, the possibility of generating at higher rates, the high-dimensional quantum systems reported in \cite{Us,Us2}, may allow their use for doing quantum communication \cite{Gisin}.

\ack

G.L. thanks Leonardo Neves for useful comments during the preparation of the manuscript. This work was supported by the Brazilian agencies CNPq, FAPEMIG and INCT-quantum information. G. L. and A. D. acknowledge the Chilean Grants Milenio ICM P06-067F, PBCT PDA-25, FONDECYT 1080383 and FONDECYT 110850555.

\end{document}